\pdfoutput=1
\documentclass[]{article}
\usepackage[utf8x]{inputenc} 
\usepackage{url} 

\usepackage{blindtext} 
\usepackage{pgfplots}
\pgfplotsset{width=7cm,compat=1.9}
\usepackage[sc]{mathpazo} 
\usepackage{lmodern}
\usepackage[T1]{fontenc} 
\usepackage{microtype} 
\usepackage[english]{babel} 
\usepackage[hmarginratio=1:1,top=32mm,columnsep=20pt]{geometry} 
\usepackage[hang, small,labelfont=bf,up,textfont=it,up]{caption} 
\usepackage{booktabs} 
\usepackage{lettrine} 
\usepackage{enumitem} 
\setlist[itemize]{noitemsep} 
\setlist[enumerate]{noitemsep} 
\usepackage{abstract} 
\usepackage[T1]{fontenc}

\usepackage{titlesec} 
\renewcommand\thesection{\Roman{section}} 
\renewcommand\thesubsection{\roman{subsection}} 
\titleformat{\section}[block]{\large\scshape\centering}{\thesection.}{1em}{} 
\titleformat{\subsection}[block]{\large}{\thesubsection.}{1em}{} 

\usepackage{fancyhdr} 
\usepackage{titling} 

\usepackage{hyperref}
\usepackage{amssymb}
\usepackage{alltt}
\usepackage{float}

\hypersetup{
	pdfauthor={Adán Sánchez de Pedro Crespo <adan@stampery.com>},
	pdftitle=Stampery Blockchain Timestamping Architecture (BTA) - Version 6,
	pdfsubject={A method for timestamping, anchoring and certification of a virtually unlimited amount of data in one or more blockchains, focusing on scalability and cost-effectiveness while ensuring existence, integrity and ownership by using cryptographic proofs that are independently verifiable by anyone in the world without disclosure of the original data and without the intervention of the certifying party.},
	pdfkeywords={blockchain, bitcoin, ethereum, timestamping, timestamp, data anchoring, bta, stampery}
}

\pretitle{\begin{center}\Huge\bfseries} 
\posttitle{\end{center}} 
\title{Stampery Blockchain Timestamping Architecture (BTA)} 
\author{%
	\textsc{Adán Sánchez de Pedro Crespo}\\
	\textsc{\small Stampery, CTO}\\
	\normalsize \href{mailto:adan@stampery.com}{adan@stampery.com} 
	\and
	\textsc{Luis Iván Cuende García} \thanks{Luis Iván---Stampery's former CTO and now one of our advisors---is credited here for being the original deviser and developer of BTA's very first version back in 2014.}\\
	\textsc{\small Aragon, CEO}\\
	\normalsize \href{mailto:luis@aragon.one}{luis@aragon.one} 
}

\date{\textbf{Version 6 -- LTS}\\Published on October 16, 2016 \\Last reviewed on November 13, 2017} 
\renewcommand{\maketitlehookd}{%
	\begin{abstract}
		\noindent A method for timestamping, anchoring and certification of a virtually unlimited amount of data in one or more blockchains, focusing on scalability and cost-effectiveness while ensuring existence, integrity and ownership by using cryptographic proofs that are independently verifiable by anyone in the world without disclosure of the original data and without the intervention of the certifying party.
	\end{abstract}
}

\begin{document}

\maketitle


\section{Introduction}

\lettrine[nindent=0.5em,lines=3]{B}{lockchains}---understood as immutable, decentralized, distributed and unedited digital ledgers---have already proved their undeniable capability to serve as a universal support for a distributed monetary system.

Since the appearance of Bitcoin\cite{bitcoin:paper} and its revolutionary approach to digital cash, there have been many attempts to develop methods allowing anyone to leverage the immutability of the blockchain technology for purposes other than the transfer of currency. 

The Bitcoin protocol itself provides an opcode\cite{wiki:opcode} called OP\_RETURN\cite{bitcoin:opreturn} that lets the participants of the network to embed arbitrary data into the transactions they broadcast to the Bitcoin blockchain since version 0.9.0 of {\em Bitcoin Core}, released in March 2014.

Since OP\_RETURN was made available, tools like Proof of Existence\cite{pof}, Stampery\cite{stampery:website} or Tierion\cite{tierion} have been making use of the Bitcoin blockchain to timestamp and verify data.

Nevertheless, the OP\_RETURN method has several limitations that are intrinsic to the very specificities of the bitcoin protocol:

\begin{itemize}
	\item In order to keep the average transaction and block size inside acceptable bounds, the maximum length for a single OP\_RETURN operation is 80 bytes.
	\item It is considered non-standard for a transaction to carry more than one OP\_RETURN output.
	\item Due to the existing block size limit, the bitcoin network can handle up to 7 transactions per second under ideal circumstances, but the actual value is even lower.
	\item The fee that must be paid to the network in order to rest assured that all transactions are accepted and validated in a reasonable time span is high and may keep rising over time as Bitcoin value fluctuates. Broadcasting a transaction containing a single OP\_RETURN with a certain guarantee that it will be written into the next block and confirmed nearly immediately can cost up to 120,000 satoshi (0.0012 Bitcoins), which at the time of the last review of this document roughly amounts to \$3.12.
\end{itemize}

In a very first approach to overcoming some of those limitations, the aforementioned timestamping services do not actually embed the whole piece of data to be timestamped. Instead, they calculate a cryptographic hash of the data, which serves as a univocal identifier that has a length between 32 and 64 bytes and therefore fits into a single bitcoin transaction. This is what we refer as \textit{data anchoring}, \textit{blockchain timestamping} or simply \textit{stamping}.

Nevertheless, there are some use cases which require bulk timestamping of hundreds or thousands of files; or timestamping the same piece of data over and over in its successive versions. This would require expending a huge amount of money in transaction fees and would quickly lead to saturation of the previously mentioned bitcoin block size limit. These consequences would obviously render the OP\_RETURN method unsuitable for such use cases.

It is worth mentioning that other Proof-of-Work blockchains like Ethereum's or Litecoin's do not offer a definitive solution for those problems. Although they may benefit from lower transaction fees at the moment, given a huge volume of stamping actions, the limitations would be roughly the sames.

This technical paper proposes a method and architecture that allows (1) carrying out scalable and cost-effective blockchain data anchoring (overcoming all the already referred limitations), (2) generating irrefutable proofs of anchoring and (3) verification of the validity of such proofs by any individual or system in the world.

\section{Blockchain Anchoring}
\label{sec:anchoring}

\lettrine[nindent=0.5em,lines=2]{B}{lockchain} anchoring or blockchain timestamping consists in leveraging the immutability and security features of public blockchain technologies to certify, ensure and prove that a document, communication or dataset existed at a certain point in time by generating a digital proof of existence.

Unlike signature-based timestamping and other digital evidentiary tools that existed before the emergence of blockchains and their consensus algorithms, blockchain anchoring does not require relying on a potentially corruptible third party as a source of truth.

Instead, blockchain anchoring relies on mathematically provable digital documents that are backed by the same consensus algorithms found in the blockchains used to perform the anchoring process. The conflicting interests of the numerous and heterogeneous participants of blockchain networks make it impossible for a majority of them to collude with the purpose of altering or tampering with the timestamps. 

Given the historic evolution of the volume of monetary value stored in form of tokens or digital coins in the main public blockchains, the chance for these to completely disappear in the next decades or even centuries is very limited, which ensures that proofs generated by blockchain anchoring methods will be available and enforceable far in the future. 

Most blockchain anchoring services make use of one-way “hash functions” to generate unique fingerprints of the data being anchored and publicly distribute those in such a manner that anyone holding the original data can verify that they match, but it is impossible for anyone else to retrieve or reconstruct the original data with the only knowledge of the fingerprint. This property guarantees that the proofs are completely free of privacy concerns and therefore they can be published safely on public blockchains.

From a technical standpoint, the stamping or anchoring itself is performed by:
\begin{enumerate}
    \item Applying a hash function (commonly SHA-256\cite{wiki:sha256}) to the data unit to be anchored, 
    \item writing the resulting hash into a blockchain transaction, 
    \item broadcasting the transaction into one or more blockchains, and finally
    \item distributing or storing the data of the blockchain transaction along the metadata of the anchored data unit.
\end{enumerate}

The time assumed to be the timestamp is the one contained in the information of the block in which the broadcasted transaction gets included, which can be easily looked up in block explorer websites and services. While the precision and accuracy of these timestamps is not comparable to that of atomic clocks, it happens to be good enough for most use cases.

Because of the increasing cost of broadcasting transactions in the main public blockchains, there are some cryptographic constructions that can be applied in order to aggregate several data units into only one hash and send them in a single blockchain transaction in a similar fashion to how transactions themselves are aggregated into every block in a blockchain. This makes the cost of the anchoring process dependent on the period between transactions that the anchoring party decide to establish rather than on the number of data units being anchored. This way, a virtually unlimited number of data units can be anchored into a single blockchain transaction, making the unitary cost of the each stamp approach asymptotically to zero.

\begin{figure}[H]
\centering
\parbox{7.1cm}{
    \[tc(n) = n \cdot fee \in \mathcal{O}(n)\]
    \[cpt(n) = \frac{tc(n)}{n} = fee \in \mathcal{O}(1)\]\[\]
    \centering
    \begin{tikzpicture}
        \begin{axis}[
            axis lines=left,
            ymax=6,
            xtick={1,3,5,7,9},
            ytick={1,3,5,7,9}
        ]
            \addplot[color=red, domain=0:6]{x};
            \addlegendentry{$tc(n)$}
            \addplot[color=blue, domain=0:6]{1};
            \addlegendentry{$cpt(n)$}
        \end{axis}
    \end{tikzpicture}
    \caption{Without aggregation, the total cost is directly proportional to the number of timestamps (linear) while the cost per timestamp always equals the transaction fee (constant).
}
\label{fig:aggregationA}}
\qquad
\begin{minipage}{7.1cm}
    \[tc(n) = fee \in \mathcal{O}(1)\]
    \[ppt(n) = \frac{tc(n)}{n} = \frac{fee}{n} \in \mathcal{O}(\frac{1}{n})\]\[\]
    \centering
    \begin{tikzpicture}
        \begin{axis}[
            axis lines=left,
            ymax=6,
            xtick={1,3,5,7,9},
            ytick={1,3,5,7,9}
        ]
            \addplot[color=red, domain=0:6]{1};
            \addlegendentry{$tc(n)$}
            \addplot[color=blue, domain=0:6, samples=50]{1/x};
            \addlegendentry{$cpt(n)$}
        \end{axis}
    \end{tikzpicture}
    \caption{With aggregation, the total cost always equals the transacion fee (constant) while the cost per timestamp is inversely proportional to the number of timestamps (rectangular hyperbola).
}
\label{fig:aggregationB}
\end{minipage}
\end{figure}

\section{Hashes aggregation}
\label{sec:aggregation}

Many single hashes corresponding to different pieces of data or files can be compiled into a single hash by building a \textit{binary tree} in which every leaf node is populated with each of the hashes while every non-leaf node is populated with the hash of the merger of its child nodes. This type of tree is commonly known as \textit{Merkle tree}\cite{wiki:merkle}.

\begin{figure}[h]
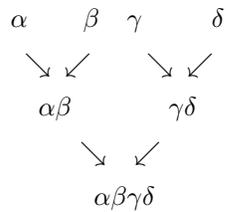

	\begin{alltt}
		\(\alpha\)    \(\beta\)  \(\gamma\)     \(\delta\)
		 \(\searrow \swarrow\)    \(\searrow \swarrow\)
		  \(\alpha\beta\)       \(\gamma\delta\)
		     \(\searrow  \swarrow\)
		      \(\alpha\beta\gamma\delta\)
	\end{alltt}
	\caption{Example of Merkle tree with 4 hashes as leaf nodes.}
\end{figure}

Merkle trees have been used for decades to perform efficient and secure verification of the contents of large data structures\cite{merkle}. Indeed, popular blockchain systems like Bitcoin, Ethereum or Litecoin rely on Merkle trees for gathering together transactions into every block of the chain.

The tip node in the tree---the hash resulting from aggregation of all the hashes in the leafs---is called \textit{Merkle root} or simply \textit{root}.

The binary hash chain\cite{wiki:hashchain} proving that a certain hash belongs to a tree---further discussed in Section~\ref{sec:proofs}---is called \textit{Merkle proof} or simply \textit{proof}.

\begin{figure}[h]
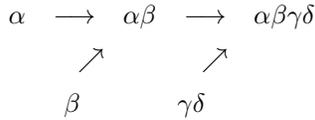

	\begin{alltt}
		\(\alpha\)  \(\longrightarrow\)  \(\alpha\beta\)  \(\longrightarrow\)  \(\alpha\beta\gamma\delta\)
		     \(\nearrow\)       \(\nearrow\)
		    \(\beta\)       \(\gamma\delta\)
	\end{alltt}
	\caption{Binary hash chain representing the Merkle proof for the $\alpha$ hash in the tree depicted in \textbf{Figure 1}. $\Pi(\alpha) = \{\beta, \gamma\delta\}$}
\end{figure}

By aggregating multiple hashes into a Merkle tree and then publishing only the root, it is possible to anchor large volumes of data in a blockchain with a single transaction, which dramatically reduces costs and avoids the aforementioned bottlenecks.

The algorithm followed to calculate the hash in each non-leaf node by joining its two child nodes is called a \textit{mixing fuction}, \textit{Merkle mixer} or simply \textit{mixer}.

A very common mixer---used by the Stampery API and Chainpoint---can be expressed in the form \textsc{SHA256($\lambda + \rho$)} and represents the result of calculating the SHA-256 hash of the concatenation of the left child $\lambda$ and the right child $\rho$. However, given that concatenation is a non-commutative binary operation (\textsc{$\lambda + \rho \not= \rho + \lambda$}), it makes necessary to distribute some additional bits of information along the Merkle proofs, being them (a) the position (index) in which each leaf hash was added, or (b) the order in which every pair of child nodes were concatenated.

To avoid the need to deal with those \textit{order bits}, previous versions of BTA used a different approach to mixing functions: \textit{commutative concatenation}.

Commutative concatenation, $\dotplus$, of two values $\lambda$ and $P$ can be easily achieved by ordering them from lowest to highest before performing the concatenation. This way, \textsc{$\lambda \dotplus \rho = \rho \dotplus \lambda$}.

Former versions of BTA proposed the use of the \textit{SHA-3 hash algorithm} as defined by FIPS-202\cite{sha3} with a 512 bit key to ensure resistance to \textit{length extension attacks} and many other theoretical vulnerabilities affecting SHA-2 and all other algorithms based on the \textit{Merkle–Damgård construction}\cite{wiki:lea}. Nevertheless, the current versions of BTA and the Stampery API use SHA-256 by default for the sake of ease of integration into a wider range of platforms and devices.

\begin{figure}[h]
\begin{alltt}
\textbf{algorithm} mixer \textbf{is}
  \textbf{input:} left hash \(\lambda\),
         right hash \(\rho\)
  \textbf{output:} parent node hash \(\pi\)
  
  \(\pi\) ← SHA256(\(\lambda + \rho\))
  \textbf{return} \(\pi\)
\end{alltt}
\caption{The non-commutable mixing function used by the Stampery API, described in pseudocode.}
\end{figure}

Due to the way that binary trees work, the root is always connected to every leaf hash by a unique path with the exception of the cases when the number of nodes in one of the levels in the tree is odd. In such cases, in order to find a root that is connected to all leaf hashes, "orphan" nodes must be (a) promoted to the next level without applying any mixing function to its value; or (b) mixed with a random string of the same length of the actual hashes. This process is called \textit{closing} a Merkle tree.

\begin{figure}[h]
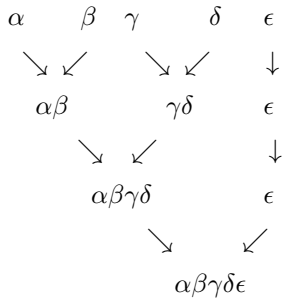

\begin{alltt}
      \(\alpha\)    \(\beta\)  \(\gamma\)     \(\delta\)   \(\epsilon\)
       \(\searrow \swarrow\)    \(\searrow \swarrow\)    \(\downarrow\)
  	     \(\alpha\beta\)       \(\gamma\delta\)     \(\epsilon\)
  	        \(\searrow  \swarrow\)        \(\downarrow\)
  	         \(\alpha\beta\gamma\delta\)        \(\epsilon\)
  	             \(\searrow     \swarrow\)
  	               \(\alpha\beta\gamma\delta\epsilon\)
\end{alltt}
\caption{Example of closing a Merkle tree with an odd number of leaf nodes: the $\epsilon$ node is promoted to its immediately upper level until it can be merged into the root.}
\end{figure}

\section{Anchorage procedures}
\label{sec:anchorage}

As explained in the introduction, Bitcoin anchoring is not perfectly suitable for those cases in which a) latency is paramount; or b) a really fast or near-immediate confirmation is needed.

In the other hand, newer blockchain projects like Ethereum have still a lower hashing power---which translates to lower certainty and security---but they are much faster when it comes to the time needed to put a hash in a block.

\begin{table}[H]
	\centering
	\caption{Average block time comparison between highest capitalization blockchains.}
	\label{fig:blocktimecomparison}
	\begin{tabular}{ll}
		\textbf{Chain} & \textbf{Average block time} \\ \hline
		Bitcoin        & $\sim$10 minutes            \\
		Ethereum       & $\sim$17 seconds            \\
		Ripple         & $\sim$4 seconds             \\
		Litecoin       & $\sim$2.5 minutes
	\end{tabular}
\end{table}

BTA is designed to allow anchoring data into more than one blockchain, leveraging the advantages of each one. A recommended strategy is combining Bitcoin and Ethereum anchoring together to get certainty and responsiveness at the same time.

Anchoring to multiple chains and different ways to do so will be further discussed in Section~\ref{sec:multiple}.

\subsection{Bitcoin anchoring}

Bitcoin anchoring is achieved by using the aforementioned OP\_RETURN opcode.

The OP\_RETURN data length limit used to be set to 40 bytes before February 2015. Then the limit was raised to 80 bytes\cite{github:opreturn}, which allowed the use of longer hashing digests.

The hashes resulting from the recommended hashing algorithm in this document---SHA-2 in its 256 bit version---are 32 bytes long, so they fit perfectly into a single OP\_RETURN instruction.

Addition of a OP\_RETURN operation to a Bitcoin transaction is made by adding a zero-value output with a {\em scriptPubKey} consisting of "\texttt{6A \(\sigma\) \(\delta\)}", where $\sigma$ is the size in bytes of the data to attach to the transaction, and $\delta$ is the data itself. Both values are encoded in hexadecimal notation.

\begin{figure}[h]
\begin{alltt}
 6A 20 9D0F5692F0A7CCDBE5554732094CFB52
       589F5D8AD762BB54B77A1978462F01C2
\end{alltt}
\caption{Sample scriptPubKey for anchoring a SHA-256 hash. Spaces and newlines must be ignored.}
\end{figure}

A relatively high fee must be paid in order to rest assured that a transaction gets successfully accepted and confirmed by the Bitcoin network in a reasonable time even under heavy load circumstances. Recommended values are between 60,000 and 120,000 satoshi.

\subsection{Ethereum anchoring}

The Ethereum protocol provides a convenient method for anchoring hashes or data into transactions. This is done by creating a \textit{message call transaction} \textsc{$T$} in which the data attribute \textsc{$T_{d}$} is set to be the Merkle root.

The protocol expects \textsc{$T_{d}$} to be an unlimited size byte array\cite{eth:paper}, so again the hashes resulting from the recommended hashing algorithm in this document---SHA-2 in its 256 bit version---fit perfectly into a single transaction.

A gas\footnote{Gas is the pricing or fee for running a transaction or contract in Ethereum.} value of 24000 szabo at a 20 Gigawei gas price is typically needed for broadcasting a message call transaction containing a SHA-256 hash.

\subsection{Prefixing}

Following Bitcoin community's convention for prefixing OP\_RETURN operations in order to identify the generating application, a prefix should be added in front of the data itself, no matter which blockchain we are anchoring to.

At the time of this writing, there is no standarized format for prefixes, so different BTA implementations may choose to use it in a different manner.

By way of illustration, Stampery's BTA implementation uses the prefix format "\texttt{53 \(\upsilon\) \(\phi\)}", where \texttt{53} represents the letter "S"---standing for "Stampery"---in ASCII format, $\upsilon$ is one byte telling the BTA version again in ASCII, and $\phi$ is a two-bytes string identifying the server that broadcasted the transaction.

\begin{figure}[h]
	\begin{alltt}
		6A 24 53 35 9B73
		      9D0F5692F0A7CCDBE5554732094CFB52
		      589F5D8AD762BB54B77A1978462F01C2
	\end{alltt}
	\caption{Same scriptPubKey used in \textbf{Figure 5}, this time with proper prefixing and values $\upsilon=5$, $\phi="9B73"$. Spaces and newlines must be ignored.}
\end{figure}

The $\phi$ value may be calculated by taking the first two bytes from the MD5 hash of the server's host name.

\section{Proof of anchorage}
\label{sec:proofs}

As mentioned before, the \textit{binary hash chain}\cite{wiki:hashchain} that proves that a certain hash belongs to a tree is called \textit{Merkle proof} or simply \textit{proof}.

The proof for a single leaf hash contains the sibling hashes that need to be sequentially concatenated and rehashed in pairs in order to reconstruct the path from the leaf to the root.

\begin{figure}[h]
\begin{alltt}
\(\Pi(\alpha) = \{\beta, \gamma\delta, \epsilon\}\)   \(\Pi(\gamma) = \{\delta, \alpha\beta, \epsilon\}\)
\(\Pi(\beta) = \{\alpha, \gamma\delta, \epsilon\}\)   \(\Pi(\epsilon) = \{\alpha\beta\gamma\delta\}\)
\(\Pi(\delta) = \{\gamma, \alpha\beta, \epsilon\}\)
\end{alltt}
\caption{Example illustrating the proofs ($\Pi$) for all the hashes in the Merkle tree from \textbf{Figure 4}. As it is noticeable, two contiguous hashes will likely have similar proofs, with the only one or two first items in the hash chain being different.}
\end{figure}

In order to allow independent individuals and systems to verify the proofs of anchorage, there are some other elements that need to be distributed along the hash chain and the Merkle index. Those are: (1) the hash of the original data piece, (2) a reference to the chain in which the root was embedded, and (3) a transaction identifier that unambiguously points to the transaction where the root was embedded.

\subsection{Proof formatting for internal use}

Stampery's approach to proof formatting is really minimalistic and concise. A BTA proof of anchorage consists of one array or list containing the following items:

\begin{itemize}
	\item $\Pi_{[0]}$: BTA version (currently \textsc{6})
	\item $\Pi_{[1]}$: Merkle index
	\item $\Pi_{[2]}$: Merkle proof (siblings list/hash chain)
	\item $\Pi_{[3]}$: Merkle root
	\item $\Pi_{[4]}$: Anchor tuple
	\begin{itemize}
		\item $\Pi_{[4][0]}$: Chain ID
		\item $\Pi_{[4][1]}$: Prefix
		\item $\Pi_{[4][2]}$: Transaction ID
	\end{itemize}
\end{itemize}

Current BTA version is 6. Prior versions used a different proof format which is not compatible with the one proposed here.

The Merkle root $\Pi_{[2]}$ and all the hashes in the proofs list $\Pi_{[1][n]}$ should be printed using uppercase. On the contrary, the transaction ID can be printed using either uppercase or lowercase.

Merkle proof lists with zero siblings can be represented by an empty brackets \textsc{[]}, empty quotation marks \textsc{""} or a \textsc{null} value.

\begin{figure}[h]
\begin{alltt}
[
 6,
 2,
 [
  "EFA4BA8F7A66BC3B\(\textellipsis\)D3C358038F5A9C27"
  "7DDE76C5E472C9AE\(\textellipsis\)B7B9567DCB3E9551"
 ],
 "1FA2DF8A8ABEA78E\(\textellipsis\)EDC4E25207D2F125",
 [
   1,
   "9B73"
   "84ba00d2cebbb4ee\(\textellipsis\)fbb06a053e4fba10"
 ]
]
\end{alltt}
\caption{A BTA proof of anchorage, represented using JSON format. All the hashes and the transaction ID have been shortened for fitting this document's layout.}
\end{figure}

\subsection{Chainpoint and OpenTimeStamps compatibility}

\begin{figure}[h]
\begin{alltt}
	
\{
  "@context": "https://w3id.org/chainpoint/v2",
  "type": "ChainpointSHA256v2",
  "targetHash": "9D0F5692F0A7CCDB\(\textellipsis\)B77A1978462F01C2",
  "proof": [
    \{
      "left": "EFA4BA8F7A66BC3B\(\textellipsis\)D3C358038F5A9C27"
    \},
    \{
      "right": "7DDE76C5E472C9AE\(\textellipsis\)B7B9567DCB3E9551"
    \}
  ],
  "anchors": [
    \{
     "type": "BTCOpReturn",
     "sourceId": "84ba00d2cebbb4ee\(\textellipsis\)fbb06a053e4fba10"
    \}
  ]
\}
\end{alltt}
\caption{The same BTA proof from Figure 8, in Chainpoint v2 format.}
\end{figure}

BTA proofs are 100\% compatible with Chainpoint\cite{chainpoint:paper}, a similar technology developed by Tierion, Inc. This means that the proofs generated by Stampery's BTA platform and the Stampery API can be easily presented in the Chainpoint format, as shown in \textbf{Figure 9}.

Similarly, BTA proofs can be also presented in OpenTimeStamps\cite{ots:site} binary format thanks to the extensibility of such proof format and protocol.

\subsection{Chain IDs}

The \textit{chain id} is an integer number representing which blockchain is the proof anchored to.

Positive values correspond to production chains---\textit{livenets}---while negative values correspond to their test counterparts---\textit{testnets}.

\begin{table}[H]
	\centering
	\caption{Livenet Chain IDs}
	\label{fig:livenetchainids}
\begin{tabular}{cl}
	\textbf{ID} & \textbf{Chain}           \\ \hline
	1           & Bitcoin livenet          \\
	2           & Ethereum Classic Livenet \\
	3           & Ethereum Fork Livenet    \\
	4           & Litecoin
\end{tabular}
\end{table}

\begin{table}[H]
	\centering
	\caption{Testnet Chain IDs}
	\label{fig:testnetchainids}
	\begin{tabular}{cl}
		\textbf{ID} & \textbf{Chain}   \\ \hline
		-1  & Bitcoin livenet          \\
		-2  & Ethereum Rinkeby testnet \\
		-3  & Ethereum Morden testnet  \\
		-4  & Litecoin testnet
	\end{tabular}
\end{table}

ID no. \textsc{0} is left unassigned forever and shall only be used for protocol testing purposes.

\section{Proofs verification}
\label{sec:verification}

Verification of Merkle proofs is performed by reconstructing the path between the original data hash and the Merkle root. This is done by applying the mixing function to the hash and the first sibling $\Pi_{[2][0]}$ in the proof, then applying the same mixer to the resulting hash and the next sibling, and continuing this process subsequently with $\Pi_{[2][n]}$ until there are no more hashes left to merge. The resulting \textit{root candidate} hash $\rho$ is expected to correspond to the Merkle root $\Pi_{[3]}$.

The order in which the hashes are concatenated before the hashing function is applied can be easily derived from the Merkle index by using this function:

\begin{figure}[h]
\begin{center}
$o$ $=$ ($\frac{\iota}{2^{n}}$) $mod$ $2$
\end{center}
\caption{Given a merkle index $\iota$ and merkle proof $\Pi_{[2]}$, this function tells the order of concatenation $o$ for any element $\Pi_{[2][n]}$ belonging to $\Pi$. $o = 1$ means left side concatenation ($Pi_{[2][n]} + h$) and $o = 0$ means right side concatenation ($h + Pi_{[2][n]}$).}
\end{figure}

\begin{figure}[h]
\begin{alltt}
\textbf{algorithm} prover \textbf{is}
  \textbf{input:} original data hash \(\eta\),
         merkle proof \(\pi\)
         merkle index \(\iota\)
  \textbf{output:} merkle root \(\rho\)
  
  \(\rho\) ← \(\eta\)
  \textbf{for} n \textbf{in} \(\pi\)
    \(\sigma ← \pi[n]\)
    \(o ← \)(\(\iota\) / \textbf{pow}(\(2\), \(n\))) \% 2
    \textbf{if}  \(o = 1\) \textbf{do}
      \(\rho = \) \textbf{mixer}(\(\rho\), \(\sigma\))
    \textbf{else do}
      \(\rho = \) \textbf{mixer}(\(\sigma\), \(\rho\))
  \textbf{return} \(\rho\)
\end{alltt}
\caption{An iterative proof verification function, described in pseudocode. The \textit{mixer} function is the one described in \textbf{Figure 3}.}
\end{figure}

In the event that the resulting hash and the Merkle root mismatched, either the proof or the original data hash were modified, corrupted or tampered with at some time after the anchoring took place. This can be used as a method for checking file integrity over time---time stamp the same file over and over and keep a history of the data hashes and proofs: if the file gets modified, the hash will change, will not match the one related to the previous proof and the date of modification will become evident.

In order to verify the validity of the proof and the anchor, it is also necessary to check if the root candidate or the Merkle root match the data payload in the transaction. This can be done by searching the transaction ID $\Pi_{[4][1]}$ in a block explorer\footnote{Some popular block explorers are Blockcypher for Bitcoin and Etherscan for Ethereum.}, reading the payload and then checking whether it contains the root as a substring---probably at the end because of prefixing.

\begin{figure}[h]
\begin{center}
$\eta = \Pi_{[2]}$\\
$\eta \in \delta$
\end{center}
\caption{The candidate root $\eta$ should match and be a substring of the data payload $\delta$ found in the anchoring transaction.}
\end{figure}

\section{Progressive proofing}
\label{sec:progressive}

Hashing is a quite intensive operation in terms of CPU usage. When dealing with big Merkle trees---with hundreds, thousands or even millions of leaf nodes---closing the tree and finding a final Merkle root is not trivial. Specifically, For any number $\nu$ of child nodes, the number of hashing operations needed to find the root is $\nu - 1$.

In addition, calculating the proof for every leaf node once the tree is closed means a lot of time wasted traversing it and picking the siblings.

Those are the reasons why we recommend to build the Merkle tree and the proofs as the hashes come and are appended to the leaf level instead of doing it at the time of closing.

Taking this into account, the logic to be carried out every time a new leaf node is added is described in \textbf{Figure 11}.

\begin{figure}[h]
\begin{alltt}
\textbf{algorithm} pusher \textbf{is}
  \textbf{input:} merkle tree \(\tau\),
         proofs list \(\pi\),
         new hash \(\eta\),
         level \(\lambda\) (default = 0)  
  \textbf{output:} merkle tree \(\tau\),
          proofs list \(\pi\)
  
  \textbf{if} \(\eta \not\in \tau[0]\) \textbf{do}
    \(\sigma ← \tau[\lambda]\) or \([]\)
    \(\sigma ← \sigma + [\eta]\)
    
    \textbf{if} \(\lambda = 0\) \textbf{do}
      \(\pi[\eta] ← []\) 
	  
    \textbf{if} \(mod(|\tau[\lambda]|)\) = 1 \textbf{do}
      \(\varsigma ← \tau[\lambda][-1]\)
      \(\varpi ← mixer(\varsigma, \eta)\)
      \(\pi ← proofer(\tau, \pi, \eta, \varsigma, \lambda)\)
      \(\{\tau, \pi\} ← pusher(\tau, \pi, \varpi, \lambda + 1)\)       
     
  \textbf{return} \(\{\tau, \pi\}\)
\end{alltt}
\caption{A recursive hash appending routine, described in pseudocode. The \textit{mixer} function is the one in \textbf{Figure 3}, and the proofer function is described in \textbf{Figure 11}.}
\end{figure}

\begin{figure}[h]
\begin{alltt}
\textbf{algorithm} proofer \textbf{is}
  \textbf{input:} merkle tree \(\tau\),
         proofs list \(\pi\),
         left hash \(\alpha\),
         right hash \(\beta\),
         level \(\lambda\) (default = 0)  
  \textbf{output:} proofs list \(\pi\)
  
  \(\phi ← \tau[0]\)
  \(\kappa ← \tau[\lambda]\)
  \(\rho ← ⌊2^{(\lambda + 1)}⌉\)
  \(\sigma ← ⌊-\rho + |\kappa| × (2^{\lambda})⌉\)
  \(\epsilon ← min((\sigma + \rho), |\phi|) - 1\)
  \(\iota ← \sigma\)
  
  \textbf{for each} hash \((\chi)\) \textbf{in} \(\phi \in [\sigma, \epsilon]\) \textbf{do}
    \textbf{if} \(\iota < ⌊\rho / 2⌉\) \textbf{do}
      \(\pi[\chi] ← \pi[\chi] + [\alpha]\)
    \textbf{else do}
      \(\pi[\chi] ← \pi]\chi] + [\beta]\)
    \(\iota ← \iota + 1\)
      
  \textbf{return} \(\pi\)
\end{alltt}
\caption{Auxiliar function for progressive proofing used by the pusher function in \textbf{Figure 11}.}
\end{figure}

\section{Anchoring to multiple chains}
\label{sec:multiple}

As mentioned before, it is specially convenient to perform anchoring to multiple blockchains in order to leverage the best of each one's feature set and also to have auxiliary anchors just in case some day one blockchain is somehow hacked or corrupted and the anchoring transaction is reverted, tampered with or simply disappears.

\subsection{Parallel anchorage}

Parallel anchorage is the simplest way to anchor Merkle roots to multiple chains.

It involves closing the Merkle tree every few minutes, calculating the root and then embedding it into multiple blockchains at the same time.

This way, the proofs for all the chains are identical with the exception of their last element---the anchor tuple $\Pi_{[4]}$---that will have different values for the chain ID $\Pi_{[3][0]}$ and transaction ID $\Pi_{[3][2]}$.

\subsection{Incremental anchorage}

Incremental anchorage is a more complex but also more powerful way to anchor Merkle roots to multiple chains.

It involves creating several different types of Merkle trees, each type corresponding to a chain we want to anchor to. Every of them will have its own \textit{lifetime}---time before tree closing---based on the average block time of the matching chain.

For example, if we were to perform incremental anchorage with Ethereum and Bitcoin, we could define the following tree types and tree lifetimes:

\begin{table}[H]
	\centering
	\label{fig:treelifetime}
	\begin{tabular}{ll}
		\textbf{Chain} & \textbf{Tree lifetime} \\ \hline
		Ethereum       & 1 minute              \\
		Bitcoin        & 10 minutes            \\
	\end{tabular}
\end{table}

A incremental anchorage procedure for those chains and lifetimes could be the following:

\begin{enumerate}
	\item Create one tree for Ethereum and one for Bitcoin.
	\item Receive all the hashes as they come and push them as leafs into Ethereum tree.
	\item After 1 minute:
	\begin{enumerate}
		\item Close Ethereum tree and get its Merkle root.
		\item Anchor the root to Ethereum blockchain.
		\item Deliver the Ethereum proofs to the requesting parties.
		\item Push the Ethereum root into the Bitcoin tree as if it were a leaf hash.
		\item Create a new Ethereum tree for the next 1-minute frame.
	\end{enumerate}
	\item Repeat steps 2 and 3 until minute 10.
	\item When minute 10 comes:
	\begin{enumerate}
		\item Close Bitcoin tree and get its Merkle root.
		\item Anchor the root to Bitcoin blockchain.
		\item Append each Bitcoin proof's siblings list $\Pi_{[2]}$ (\textit{proof head}) to the end of all the Ethereum proofs' siblings list $\Pi′_{[2]}$ (\textit{proof tail}) having a root $\Pi′_{[3]}$ that matches the original data hash from the Bitcoin proof.
		\item Deliver the whole "merged" proofs to the requesting parties.
		\item Delete all the old Ethereum trees and create a new Bitcoin tree for the next 10-minutes frame.
	\end{enumerate}
	\item Start over the whole process.
\end{enumerate}

Because of the way the Merkle proofs are generated and merged together, the resulting Ethereum proof will be a subset of the Bitcoin proof, but both of them will still be perfectly valid separately.

\section{Distributed BTA (DBTA)}
\label{sec:distributed}

It is easy to imagine that a computer handling a Merkle tree with thousands or millions of nodes in its leaf level will need a huge amount of RAM memory. Furthermore, if the computer runs out of memory while building a Merkle tree, it will never be able to close it and all the hashes will be simply lost.

This kind of bottlenecks and situations can be easily overcome by using a Distributed BTA implementation as described hereafter.

DBTA is a computation cluster formed by a number $\nu$ of BTA instances called \textit{nodes}.

A messaging queue system is used to distribute messages and balance workload between all the nodes.

In addition, a P2P protocol in conjunction to a basic consensus algorithm are used to reduce the number of Bitcoin transactions made by the cluster.

\subsection{The cluster}

We analyzed many different technologies and platforms when we built Stampery's own DBTA cluster, and finally opted for Erlang/OTP.

Erlang\cite{erlang:paper} is a programming language and virtual machine used to build massively scalable soft real-time systems with requirements on high availability. Some of its uses are in telecoms, banking, e-commerce, computer telephony and instant messaging. Erlang's runtime system has built-in support for concurrency, distribution and fault tolerance.

OTP is a set of Erlang libraries and design principles providing middle-ware to develop these systems. It includes its own distributed database, applications to interface towards other languages, debugging and release handling tools.

For convenience, we wrote our BTA implementation using Elixir\cite{elixir:site}, a dynamic, functional language designed for building scalable and maintainable applications.

\subsection{The messaging queue}

The messaging queue will fulfill a variety purposes, most of them related to the incremental anchoring procedure as described by Section~\ref{sec:progressive}:
\begin{itemize}
	\item Load balancing the distribution of incoming hashes to the nodes.
	\item Keeping the hashes in memory until they have been successfully anchored to all chains and the corresponding proofs have been delivered.
	\item Routing the Ethereum proofs back to their requesting parties.
	\item Load balancing the distribution of Ethereum roots from one node to another, so that they all receive an even amount of Ethereum roots to use as leaf hashes in their Bitcoin trees.
	\item Routing the Bitcoin proof heads to the nodes holding the matching proof tails.
	\item Routing the whole "merged" proofs to the requesting parties.
	\item Re-delivering hashes and roots to different nodes in case the ones that were assigned originally fail to perform their functions.
\end{itemize}

Therefore, the messaging queue system that we need must fulfill all those load balancing, routing and re-delivering requirements.

That is the case with RabbitMQ, the most notable implementation of the AMQP protocol\cite{amqp:paper}. We have been successfully using it in Stampery for DBTA since 2015 and it has proven to be the perfect fit for the set of features previously described.

To implement DBTA in RabbitMQ, the following queues are needed:

\begin{table}[h]
	\centering
	\begin{tabular}{c}
		\textbf{Queue name} \\ \hline
		eth                 \\
		btc                
	\end{tabular}
\end{table}

In addition, it will be necessary to define the following exchanges\footnote{Exchanges are AMQP entities where messages are sent. Exchanges take a message and route it into zero or more queues. The routing algorithm used depends on the exchange type and rules called bindings.}:

\begin{table}[h]
	\centering
	\begin{tabular}{cc}
		\textbf{Exchange name} & \textbf{Type} \\ \hline
		eth                    & direct		   \\
		btc                    & direct        \\
		proofs                 & direct        \\
	\end{tabular}
\end{table}

When a hash is received by a DBTA implementation, it must be published to the \textsc{eth} exchange, which will route it to a different DBTA node every time\footnote{A direct exchange delivers messages to queues based on the message routing key. Direct exchanges are often used to distribute tasks between multiple workers (instances of the same application) in a round robin manner.)}. At the same time, a binding will be created in the proofs exchange, using the hash as the routing key and pointing to a user-specific proof delivery queue, named in the form of a user identifier plus the $-clnt$ suffix.

The node will then push the hash into its latest Ethereum tree and will wait 1 minute before closing it. During that time, it will predictably receive approximately $\frac{\eta}{\nu}$ hashes, where $\eta$ is the number of hashes sent to the exchange during the 1-minute frame, and $\nu$ is the number of DBTA nodes consuming the \textsc{eth} queue.

Once the 1-minute frame is over, all nodes will close their Ethereum trees, get their roots and publish them to the \textsc{btc} exchange. Such exchange will then route every root to a different DBTA node. At the same time, another binding will be created in the proofs exchange, using the root as the routing key and pointing to the same user-specific proof delivery queue, named as already stated.

The nodes receiving the roots from the \textsc{btc} exchange will treat them as if they were hashes, push them into their Bitcoin trees and wait for 10 minutes before closing the trees. During that time, each node will predictably receive up to a maximum of $10$ roots.

Once the 10-minutes frame is over, all nodes must close their Bitcoin trees and get their final Merkle root.

One important advantage of using AMQP is that if for some reason one or more nodes go offline or gets isolated from the rest of the architecture---in case of network failure, system maintenance, manual reboot, etc--it will collect the unacknowledged messages previously sent to such node(s) and reassign them to a different set of live nodes.

\subsection{Cluster leadership}

At this point, if every DBTA node were to broadcast its own Bitcoin transaction, we would be incurring in a daily cost of up to $24\times6\times\nu\times\beta$ USD\footnote{At the time of the last review of this document, this corresponds to $\nu\times$225 USD per day.}, where $\nu$ is the number of nodes in the DBTA cluster, and $\beta$ is the price of a Bitcoin transaction in US Dollars.

Because we are focusing on scalability and cost-effectiveness, we have figured out a way to keep the number of Bitcoin transactions steady so it is always the same regardless of the number of nodes in the DBTA cluster.

The trick consists of choosing a random \textit{leader} node for every 10-minutes frame and letting it join all the roots into one and make a single transaction.

In Stampery's DBTA, this is made possible by Erlang's bundled distributed computing system\cite{erlang:distributed}.

All nodes must share the same magic cookie\footnote{When an Erlang node tries to connect to another node, the magic cookies are compared. If they do not match, the connected node rejects the connection.}, so that they will be able to connect to each other automatically as soon as their EPMD\footnote{Erlang Port Mapping Daemon} is set to discover nodes in the right domain names.

All nodes must have their clocks synchronized with one another with a maximum lead or lag that needs to be less than a half the duration of the tree lifetime corresponding to the fastest blockchain being used. This amount of time is called \textit{consensus time}.

The time frames are fixed in time so that they can be synchronized across all nodes without need for signaling. For example, the Ethereum 1-minute frames will start at 00:01:00, 00:02:00, 00:03:00, etc.; while Bitcoin 10-minutes frames will take place at 00:10:00, 00:20:00, 00:30:00 and so on.

Every time a frame expires, the nodes make a list of their peers. Then they take the peers' names, append the Epoch time\footnote{Epoch time is a system for describing instants in time, defined as the number of seconds that have elapsed since 00:00:00 (UTC), Thursday, 1 January 1970, not counting leap seconds.} for the starting second of the just finished frame, and hash the resulting string. Finally, the list of hashes is sorted alphabetically and the first one in the list is considered to be the leader.

Once the leader node has been chosen, the other nodes will report their Bitcoin roots to it, which will then build a final Merkle tree joining its own Bitcoin root and all the ones from its peers. After consensus time, it should close the tree, get the final Merkle root and anchor it to the Bitcoin blockchain.

The last thing the leader needs to do is reporting the Bitcoin transaction ID and the final proof heads to each of its peers, so that they can join them to their own Bitcoin proofs and report their proof heads back to the nodes holding the proof tails.

If the leader node failed to fulfill its duty in twice consensus time, the runner-up node in the list will take over and be the new leader for the unconfirmed frame. If it failed too, it will be the turn of the third node in the list, and so forth.

\subsection{Proofs delivery}

When an Ethereum proof is ready to be delivered to the requesting party, it just needs to be published to the \textsc{proofs} exchange using the original data hash as the routing key.

Likewise, when a Bitcoin proof is ready to be delivered to the requesting party, it just needs to be published to the same \textsc{proofs} exchange, this time using the Ethereum root as the routing key.

Due to the way in which AMQP direct exchanges work, both the Ethereum and Bitcoin proofs will be routed into the right proof delivery queue thanks to the routing keys and the previously created bindings.

\newpage
\phantomsection
\addcontentsline{toc}{section}{References}
\bibliographystyle{ieeetr}
\bibliography{main}

\begin{thebibliography}{10}

\bibitem{bitcoin:paper}
S.~Nakamoto, ``Bitcoin: A peer-to-peer electronic cash system,'' 2009.
\newblock \url{https://bitcoin.org/bitcoin.pdf}.

\bibitem{wiki:opcode}
Wikipedia, ``Opcode --- wikipedia{,} the free encyclopedia.''
\newblock
  \url{https://en.wikipedia.org/w/index.php?title=Opcode&oldid=728968013}.

\bibitem{bitcoin:opreturn}
{Bitcoin Wiki}, ``Op\_return --- bitcoin wiki.''
\newblock \url{https://en.bitcoin.it/w/index.php?title=OP_RETURN&oldid=60872}.

\bibitem{pof}
M.~Aráoz, ``What is proof of existence?,'' 2014.
\newblock \url{https://proofofexistence.com/about}.

\bibitem{stampery:website}
Stampery, ``21st century notarization,'' 2014.
\newblock \url{https://stampery.com}.

\bibitem{tierion}
Tierion, ``Blockchain proof engine,'' 2014.
\newblock \url{https://tierion.com}.

\bibitem{wiki:sha256}
{Wikipedia contributors}, ``{SHA-2} --- wikipedia{,} the free encyclopedia.''
\newblock \url{https://en.wikipedia.org/wiki/SHA-2}.

\bibitem{wiki:merkle}
Wikipedia, ``Merkle tree --- wikipedia{,} the free encyclopedia.''
\newblock
  \url{https://en.wikipedia.org/w/index.php?title=Merkle_tree&oldid=734149217}.

\bibitem{merkle}
R.~C. Merkle, {\em A Digital Signature Based on a Conventional Encryption
  Function}.
\newblock CRYPTO '87. Lecture Notes in Computer Science. 293.
\newblock p. 396.

\bibitem{wiki:hashchain}
Wikipedia, ``Hash chain --- wikipedia{,} the free encyclopedia.''
\newblock
  \url{https://en.wikipedia.org/w/index.php?title=Hash_chain&oldid=731737984}.

\bibitem{sha3}
{U.S. National Institute of Standards and Technology}, ``Sha-3 standard:
  Permutation-based hash and extendable-output functions.'' Federal Information
  Processing Standards Publication no. 202, 2015.
\newblock \url{https://dx.doi.org/10.6028/NIST.FIPS.202}.

\bibitem{wiki:lea}
Wikipedia, ``Length extension attack --- wikipedia{,} the free encyclopedia.''
\newblock
  \url{https://en.wikipedia.org/w/index.php?title=Length_extension_attack&oldid=733680211}.

\bibitem{github:opreturn}
F.~Charlon, ``Change the default maximum op\_return size to 80 bytes.''
\newblock \url{https://github.com/bitcoin/bitcoin/pull/5286}.

\bibitem{eth:paper}
{Dr. Gavin Wood}, ``Ethereum: A secure decentralised generalised transaction
  ledger - homestead revision,'' 2016.
\newblock \url{http://gavwood.com/paper.pdf}.

\bibitem{chainpoint:paper}
{Wayne Vaughan, Jason Bukowski, Shawn Wilkinson}, ``Chainpoint: A scalable
  protocol for anchoring data in the blockchain and generating blockchain
  receipts.,'' 2016.
\newblock \url{https://tierion.com/chainpoint}.

\bibitem{ots:site}
{Peter Todd, Ricardo Casatta, Luca Vaccaro, Andrew Poelstra, et. al.}, ``Open
  timestamps: A timestamping proof standard.,'' 2017.
\newblock \url{https://opentimestamps.org/}.

\bibitem{erlang:paper}
{Joe Armstrong, Bjarne Däcker, Thomas Lindgren, Håkan Millroth, et al.},
  ``Open-source erlang - white paper,'' 2000.
\newblock \url{http://www1.erlang.org/white_paper.html}.

\bibitem{elixir:site}
J.~Valim, ``Elixir,'' 2012.
\newblock \url{http://elixir-lang.org/}.

\bibitem{amqp:paper}
{Sanjay Aiyagari, Matthew Arrot, Mark Atwell, Jason Brome, Alan Conway, Robert
  Greig, Pieter Hintjens, John O'Hara, Martin Ritchie, Shahrokh Sadjadi, Rafael
  Schloming, Steven Shaw, Gordon Sim, Martin Sustrik, Carl Trieloff, Kim van
  der Riet and Steve Vinoski}, ``Amqp advanced message queuing protocol -
  protocol specification,'' 2006.
\newblock \url{https://www.rabbitmq.com/resources/specs/amqp0-9.pdf}.

\bibitem{erlang:distributed}
{Ericcson AB}, ``Distributed erlang,'' 2003.
\newblock \url{http://erlang.org/doc/reference_manual/distributed.html}.

\end{thebibliography}
	
\end{document}